\documentclass[preprint,sort&compress]{elsarticle}
\usepackage{graphicx}
\usepackage{bm}
\usepackage{fullpage}

\begin{document}

\title{Universal fractal scaling of self-organized networks}

\author[rad]{Paul J. Laurienti}
\author[bme]{Karen E. Joyce}
\author[bme]{Qawi K. Telesford}
\author[rad]{Jonathan H. Burdette}
\author[rad,bs]{Satoru Hayasaka\corref{cor1}}
\ead{shayasak@wfubmc.edu}
\cortext[cor1]{Corresponding author}

\address[rad]{Department of Radiology,
Wake Forest University Health Sciences,
Winston--Salem, North Carolina, 27157, USA
}
\address[bme]{Department of Biomedical Engineering,
Wake Forest University Health Sciences,
Winston--Salem, North Carolina, 27157, USA
}
\address[bs]{Department of Biostatistical Sciences,
Wake Forest University Health Sciences,
Winston--Salem, North Carolina, 27157, USA
}

\date{\today}

\begin{abstract}
There is an abundance of literature on complex networks describing a variety of relationships among units in social, biological, and technological systems. Such networks, consisting of interconnected nodes, are often self-organized, naturally emerging without any overarching designs on topological structure yet enabling efficient interactions among nodes.  Here we show that the number of nodes and the density of connections in such self-organized networks exhibit a power law relationship. We examined the size and connection density of 47 self-organizing networks of various biological, social, and technological origins, and found that the size-density relationship follows a fractal relationship spanning over 6 orders of magnitude. This finding indicates that there is an optimal connection density in self-organized networks following fractal scaling regardless of their sizes.
\end{abstract}

\maketitle

%
%
\section{Introduction}
There has been considerable interest in the organization of complex networks since the descriptions of small-world \cite{Watts:1998} and scale-free \cite{Barabasi:1999} networks at the end of the 1990Õs. Of particular interest are naturally occurring complex networks based on self-organizing principles \cite{Barabasi:1999}. In particular, self-organized processes have been shown to exhibit some scale-free and fractal behaviors \cite{Barabasi:1999,Bak:1987}.  Barab\'asi and colleagues demonstrated that scale-free degree distributions in many self-organized networks \cite{Barabasi:1999,Jeong:2000,Albert:2000,Barabasi:2003}, which has sparked a great debate \cite{Amaral:2000,Keller:2005,Gisiger:2001} on the actual existence of scale-free behavior in naturally occurring networks. Although the degree distributions of many networks were initially considered to follow power law distributions \cite{Gisiger:2001,Clauset:2008,Faloutsos:1999,Mitzenmacher:2006,Newman:2005}, severe truncation has often been observed \cite{Mossa:2002}. Nevertheless, it is intriguing that self-organized networks can exhibit scale-free degree distributions, and this has led scientists to the search for universality within self-organized systems. 

The literature on network organization encompasses a broad range of disciplines and disparate types of networks. The literature boasts networks that range from email communications to protein interactions to word frequencies in texts. The number of nodes and the density of connections in these networks span multiple orders of magnitude, complicating comparisons of metrics extracted from various studies. One common characteristic, however, is that the majority of them are self-organized--from social to technological to biological networks, the interactions between the nodes were not predetermined by a top-down blueprint design. Despite this lack of top-down design, some self-organized networks exhibit interesting characteristics associated with the connection density and the network size. For example, utility networks from different countries in Europe tend to have similar average degrees although these networks tremendously differ in size \cite{Carvalho:2009,Sole:2008}. Another interesting observation is that the density of connections in social networks seems to decay non-linearly as the size of the network increases \cite{Friedkin:1981}. These observations suggest that there may be an underlying relationship between the size (the number of nodes, $N$) and the connection density (the ratio of the number of existing edges to the number of all possible connections, $d$) in these network. Although these findings are based on particular types of networks, it is possible that this relationship is universal to all types of self-organized networks. Thus, in this work, we examined any universal relationship between network size  and connection density, as well as the average node degree, across various types of systems.

%
%
\section{Methods and Materials}
Network parameters from 47 unique networks were collected from the literature or publicly available databases. Table \ref{Table1} lists $N$ and $d$, as well as the average node degree $K$ and the total number of edges $m$ from these networks. Although $d$ and/or $K$ have been reported in some of these networks, these metrics are recalculated based on  $N$ and $m$ for consistency. Namely, we use the formulae $d=2m/N(N-1)$ and $K=2m/N$. Although some of the networks are directed networks, we use the formulae for undirected networks in order to focus on the density of connections regardless of their directions. Note that, from these formulae, the relationship between $K$ and $d$ can be expressed as $d=K/(N-1)$. If $N$ is sufficiently large, $d\simeq K/N$. 

In this data set, the relationship between $N$ and $d$ was examined. In particular, $d$ was expressed as an exponent function of $N$ (i.e., $d \propto N^{-\gamma}$) and the exponent describing the power-law relationship $\gamma$ was determined. Moreover, the relationship between $N$ and $K$ was also examined in a similar manner. If the observations by \cite{Carvalho:2009,Sole:2008} extends to other types of networks, $K$ should remain approximately constant over the range of $N$.

\begin{table}
\begin{tabular}{ll|ccccl}
Network classes	&	Networks		&	$N$	&	$m$	&	$d$	&	$K$	&	References	\\
\hline
\hline
Biological	&	C. Elegans metabolic		&	453	&	2033	&	1.986$\times 10^{-2}$	&	9.0	&	\cite{Duch:2005}	\\
	&	C. Elegans neural		&	277	&	1918	&	5.018$\times 10^{-2}$	&	13.8	&	\cite{Humphries:2008,Kaiser:2006}	\\
	&	E. Coli reaction		&	315	&	8915	&	0.180	&	56.6	&	\cite{Humphries:2008,Wagner:2001}	\\
	&	E. Coli substrate		&	282	&	1036	&	2.615$\times 10^{-2}$	&	7.3	&	\cite{Humphries:2008,Wagner:2001}	\\
	&	Freshwater food web	$\ddagger$	&	92	&	997	&	0.238	&	21.7	&	\cite{Newman:2003b,Martinez:1991}	\\
	&	Functional cortical connectivity		&	90	&	405	&	0.101	&	9.0	&	\cite{Humphries:2008,Achard:2006}	\\
	&	Macaque cortex		&	95	&	1522	&	0.341	&	32.0	&	\cite{Humphries:2008,Kaiser:2006}	\\
	&	Marine food web	$\ddagger$	&	135	&	598	&	6.611$\times 10^{-2}$	&	8.9	&	\cite{Newman:2003b,Huxham:1996}	\\
	&	Metabolic network		&	765	&	3686	&	1.261$\times 10^{-2}$	&	9.6	&	\cite{Newman:2003b,Jeong:2000}	\\
	&	Neural network	$\ddagger$	&	307	&	2359	&	5.022$\times 10^{-2}$	&	15.4	&	\cite{Newman:2003b}	\\
	&	Yeast protein interactions		&	2115	&	2240	&	1.002$\times 10^{-3}$	&	2.1	&	\cite{Newman:2003b,Jeong:2001}	\\
\hline
Information	&	Altavista	$\ddagger$	&	2.035$\times 10^8$	&	2.130$\times 10^9$	&	1.028$\times 10^{-7}$	&	20.9	&	\cite{Newman:2003b}	\\
	&	Book purchases		&	105	&	441	&	8.077$\times 10^{-2}$	&	8.4	&	\cite{Humphries:2008}	\\
	&	Citation	$\ddagger$	&	783339	&	6.716$\times 10^6$	&	2.189$\times 10^{-5}$	&	17.1	&	\cite{Newman:2003b}	\\
	&	RogetÕs thesaurus	$\ddagger$	&	1022	&	5103	&	9.781$\times 10^{-3}$	&	10.0	&	\cite{Knuth:1993}	\\
	&	Word adjacency		&	112	&	425	&	6.837$\times 10^{-2}$	&	7.6	&	\cite{Humphries:2008,Newman:2006a}	\\
	&	Word co-occurrence		&	460902	&	1.70$\times 10^7$	&	1.601$\times 10^{-4}$	&	73.8	&	\cite{Newman:2003b}	\\
	&	WWW nd.edu	$\ddagger$	&	325729	&	1.470$\times 10^6$	&	2.770$\times 10^{-5}$	&	9.0	&	\cite{Albert:1999}	\\
\hline
Social	&	Biology co-authorship		&	1.520$\times 10^6$	&	1.180$\times 10^7$&	1.021$\times 10^{-5}$	&	15.5	&	\cite{Newman:2003b,Newman:2001a}	\\
	&	Company directors		&	7673	&	55392	&	1.882$\times 10^{-3}$	&	14.4	&	\cite{Newman:2003b,Newman:2001b}	\\
	&	Dolphins		&	62	&	159	&	8.408$\times 10^{-2}$	&	5.1	&	\cite{Lusseau:2003,Humphries:2008}	\\
	&	Email URV		&	1133	&	5452	&	8.502$\times 10^{-3}$	&	9.6	&	\cite{Guimera:2003}	\\
	&	Email messages	$\ddagger$	&	59912	&	86300	&	4.809$\times 10^{-5}$	&	2.9	&	\cite{Newman:2003b,Ebel:2002}	\\
	&	Email address book	$\ddagger$	&	16881	&	57029	&	4.003$\times 10^{-4}$	&	6.8	&	\cite{Newman:2003b,Newman:2002}	\\
	&	Film actors		&	449913	&	2.552$\times 10^7$	&	2.521$\times 10^{-4}$	&	113.4	&	\cite{Newman:2003b}	\\
	&	Football		&	115	&	613	&	9.352$\times 10^{-2}$	&	10.7	&	\cite{Girvan:2002}	\\
	&	German directors		&	4185	&	30438	&	3.477$\times 10^{-3}$	&	14.5	&	\cite{Humphries:2008,Conyon:2006}	\\
	&	Jazz		&	198	&	2742	&	0.141	&	27.7	&	\cite{Gleiser:2003}	\\
	&	Karate		&	34	&	78	&	0.139	&	4.6	&	\cite{Zachary:1977}	\\
	&	Math co-authorship		&	253339	&	496489	&	1.547$\times 10^{-5}$	&	3.9	&	\cite{Newman:2003b,Castro:1999}	\\
	&	Newspaper article co-occurrence		&	459	&	1422	&	1.353$\times 10^{-2}$	&	6.2	&	\cite{Humphries:2008,Ozgur:2004}	\\
	&	Physics co-authorship		&	52909	&	245300	&	1.753$\times 10^{-4}$	&	9.3	&	\cite{Newman:2003b,Newman:2001a}	\\
	&	Student relationships		&	573	&	477	&	2.911$\times 10^{-3}$	&	1.7	&	\cite{Newman:2003b,Bearman:2004}	\\
	&	Telephone calls	$\ddagger$	&	4.7$\times 10^7$&	8.0$\times 10^7$	&	7.243$\times 10^{-8}$	&	3.4	&	\cite{Newman:2003b}	\\
	&	UK directors		&	8850	&	39741	&	1.015$\times 10^{-3}$	&	9.0	&	\cite{Humphries:2008,Conyon:2006}	\\
	&	US directors		&	11057	&	74414	&	1.217$\times 10^{-3}$	&	13.5	&	\cite{Humphries:2008,Conyon:2006}	\\
\hline
Technological	&	Electronic circuits		&	24097	&	53248	&	1.834$\times 10^{-4}$	&	4.4	&	\cite{Newman:2003b,Cancho:2001}	\\
	&	Internet (1998)		&	10697	&	31992	&	5.592$\times 10^{-4}$	&	6.0	&	\cite{Newman:2003b,Faloutsos:1999}	\\
	&	Internet (2006)		&	22963	&	48436	&	1.837$\times 10^{-4}$	&	4.2	&	\cite{Newman:2006b}	\\
	&	Peer-to-peer network		&	880	&	1296	&	3.351$\times 10^{-3}$	&	2.9	&	\cite{Newman:2003b,Adamic:2001}	\\
	&	Power grid (EU)		&	2783&	3762	&	9.718$\times 10^{-4}$	&	2.7	&	\cite{Sole:2008}	\\
	&	Power grid (US)		&	4941	&	6594	&	5.403$\times 10^{-4}$	&	2.7	&	\cite{Newman:2003b,Watts:1998}	\\
	&	Software classes	$\ddagger$	&	1377	&	2213	&	2.336$\times 10^{-3}$	&	3.2	&	\cite{Newman:2003b,Valverde:2002}	\\
	&	Software packages	$\ddagger$	&	1439	&	1723	&	1.665$\times 10^{-3}$	&	2.4	&	\cite{Newman:2003b,Newman:2003a}	\\
	&	Train routes		&	587	&	19603	&	0.114	&	66.8	&	\cite{Newman:2003b,Sen:2003}	\\
	&	Trans-European gas network		&	24010	&	25554	&	8.866$\times 10^{-5}$	&	2.1	&	\cite{Carvalho:2009}	\\
	&	US airlines		&	332	&	2126	&	3.869$\times 10^{-2}$	&	12.8	&	\cite{Batagelj:2006}	\\
\hline
\hline
\end{tabular}
\caption{Networks considered in the size-density relationship analysis. Parameters culled from the literature are network size ($N$) and the number of edges ($m$). The network density ($d$) and the mean degree ($K$) are calculated based on $N$ and $m$. Networks marked with $\ddagger$ contain directed connections. All other networks contain undirected connections. Note that the functional cortical connectivity network was generated by applying a threshold to a correlation matrix, yielding a network that had the edge density of approximately 10\%. }
\label{Table1}
\end{table}

\begin{figure}[h]
\includegraphics{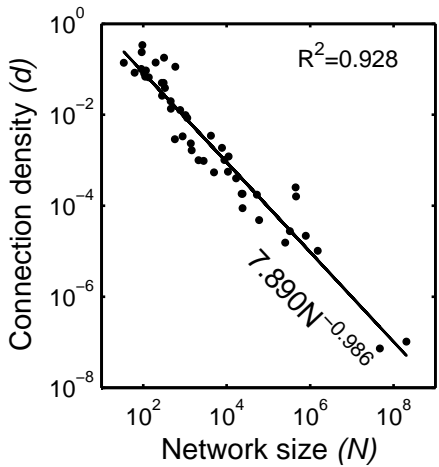}
\caption{Log-log plot of the relationship between the number of nodes in a network (network size, $N$) and the density of connections ($d$). Each point represents a different network based on the previous literature. The fit shows a power-law relationship that spans more than 6 orders of magnitude with an exponent of -0.986 consistent with a scale-free fractal behavior.}
\label{fig:dvsN}
\end{figure}

%
%
\section{Results}
When the network size $N$ and the connection density $d$ are plotted on a log-log plot (Figure \ref{fig:dvsN}), there is an obvious linear relationship between the variables. The fit to the data ($d = 7.890N^{-0.986}$) reveals a power law relationship between the size and density of the networks. The scaling exponent approaches negative one ($-1$), indicating that the relationship is fractal in nature with $1/f$ properties. Despite the wide variety of networks, there is a pronounced power law relationship between the size and the density covering more than 6 orders of magnitude. The fit to the data is very strong ($R^2 = 0.928$ on $\log_{10}$ transformed data), and there is no indication of truncation at the very large network sizes. It can be seen from Figure \ref{fig:dvsN} that there are two extraordinarily large networks included in the analysis. These networks demonstrate that there is no truncation of the relationship at the extreme values. Even when these networks are removed, the correlation remains very strong ($R^2 = 0.893$) and the exponent is $-0.978$. Thus, these two points are not unduly influencing the analysis.

\begin{figure}[h]
\includegraphics{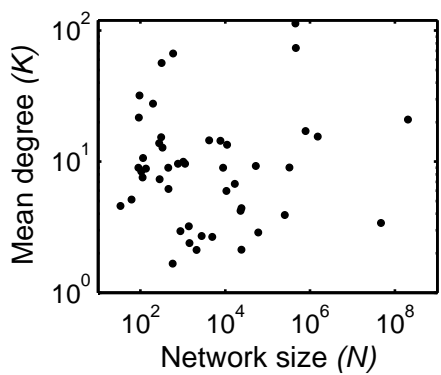}
\caption{A scatter plot of the mean degree $K$ of various networks plotted against their network size $N$. Surprisingly, $K$ does not change systematically over 6 orders of magnitude of $N$.}
\label{fig:KvsN}
\end{figure}

\begin{figure}[h]
\includegraphics{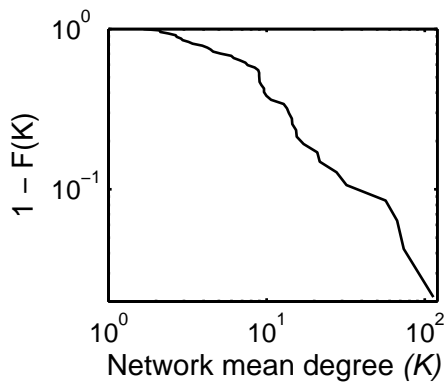}
\caption{The complementary cumulative distribution ($1-F(K)$) of the mean degree $K$. The distribution exhibits a profile of a long-tail distribution despite the limited number of observations (47 networks).}
\label{fig:DistK}
\end{figure}

When $N$ is sufficiently large, a consequence of a power-law relationship $d\propto N^{-1}$ is that $K$ does not depend on $N$. This stems from the relationship $d\simeq K/N$, which can be rewritten as $K\simeq dN = cN^{-1} N = c$ where $c$ is a constant. Since our observation above indicates a power-law relationship between $d$ and $N$ with the exponent approximately $-1$, the scatter plot of $K$ and $N$ does not indicate any association between them (see Figure \ref{fig:KvsN}). In other words, a large network size in terms of $N$ is not necessarily associated with large $K$. It is interesting to note, in Figure \ref{fig:KvsN}, that there seems to be a small number of networks with unusually large $K$ compared to the other networks. This is likely a consequence of the mean degree $K$ having a long-tail distribution, as seen in its cumulative distribution plot in Figure \ref{fig:DistK}. In this type of distributions, outliers such as $K>50$ are likely to occur while the vast majority of $K$ is reasonably small and similar. These outliers seem to occur over the range of $N$, indicating that such outliers occur randomly without any systematic dependance on $N$.

%
%
\section{Discussion}
The findings reported here demonstrate a universal relationship in self-organized networks such that the network size dictates the density. The fractal behavior observed is of particular interest because it indicates that self-organized networks are critically organized. The number of connections within each network is scaled to the size of the network, and this universal behavior likely represents an optimal organization that ensures maximal capacity at a minimal cost. Furthermore, the critical organization would indicate that a density reduction would decrease the communication capabilities of the system. Interestingly, this relationship maintains the mean degree $K$ approximately constant across different network sizes. A similar finding has been reported on the mean degree of the gas and power networks from European countries despite the large disparity in the network size \cite{Carvalho:2009,Sole:2008}. Our findings further generalizes the constant mean degree $K$ in a variety of network types. It should be noted that the relationship $d\propto N^{-1}$ is not expected from the relationship $d\simeq K/N$ alone, as $K$ could also depend on $N$ rather than being constant. To the best of our knowledge, this work is the first to demonstrate the power-law relationship between $d$ and $N$, and consequently $K$ being almost constant over 6 orders of magnitude of $N$. 

The relationship $d\propto N^{-1}$ is not surprising if one considers the emergence of a giant component in an Erd\H{o}s-R\'{e}nyi (ER)-type random graph. In an ER random graph, a giant component, or a large cluster of nodes comprising a large proportion of a network, emerges if the connection density $d$ is above the percolation threshold $p_c = 1/N$ \cite{Barabasi:1999a,Strogatz:2001}. At $d=p_c$, the corresponding value of $K$ is unity, but the observed $K$'s in real networks are always larger than unity as seen in Table \ref{Table1}. There are some possible explanations for this. First, most of the networks listed in Table \ref{Table1} are not formed in the same way as an ER random graph, but in much different mechanisms resulting in long-tail degree distributions \cite{Barabasi:1999,Mossa:2002}. In such networks, the phase transition for formation of a giant component occurs at a different threshold than that of the ER random graphs \cite{Newman:2002a}. Secondly, the actual connection density $d$ in the observed networks may be elevated compared to the percolation threshold $p_c$, resulting in $K$ being larger than unity. This may be because the networks in Table \ref{Table1} may only represent the giant component of a much larger network which includes many small isolated clusters. In other words, only a portion of the network that happened to form a connected cluster is observed, and the connection density in such a cluster may be slightly elevated compared to $p_c$ just because all the nodes are connected to that cluster.

It is true that one could artificially generate networks that do not exhibit the size-density relationship found above. In fact, the literature contains such artificially generated networks that do not lie near our plotted line. However, such artificially created networks probably do not have real world relevance. We show here the scale-free relationship between network size and connection density in real networks from such diverse origins, supporting the notion of a universal law for network organization.

While replication of these findings from additional networks will be important, there are a number of practical implications of these findings. First, the construction of networks is inherently limited by the sampling procedure used to identify nodes and links. If a self-organized network is found to disobey this relationship, one should seriously consider that there was a bias in the sampling of the network structure. Second, when building artificial networks to be compared to naturally occurring systems, the size-density relationship should roughly follow the $1/f$ relationship. For example, in studies of functional brain networks, cross-correlation matrices of nodal time series are often thresholded to identify links between nodes \cite{Bullmore:2009}. The optimal threshold to be applied is not known, and the typical solution is to utilize multiple thresholds \cite{Hayasaka:2010} producing networks with various densities. Based on the findings presented here, an optimal threshold can be easily determined, resulting in a network following the $1/f$ size-density relationship. Finally, engineered networks for practical applications may realize an optimal cost-benefit trade-off by ensuring that the density of connections is appropriate for the network size.

\section{Conclusion}
We show an important, apparently universal feature of self-organized networks: fractal scaling of size and density of connections. This fractal scaling is independent of network types, as the analysis spanned a wide gamut of networks, including biological, information, social, and technological. Thus, it appears that there is an underlying principle to organizing these self-emergent networks, a principle that probably ensures optimal network functioning.

%
%

%
%
\section*{Acknowledgements}
This work was supported in part by the National Institute of Neurological Disorders and Stroke (NS070917, NS056272, and NS039426-09S1) and the Translational Science Institute of Wake Forest University (TSI-K12).

%
%
\bibliographystyle{elsarticle-num}
\bibliography{FractalScaling}

\begin{thebibliography}{10}
\expandafter\ifx\csname url\endcsname\relax
  \def\url#1{\texttt{#1}}\fi
\expandafter\ifx\csname urlprefix\endcsname\relax\def\urlprefix{URL }\fi
\expandafter\ifx\csname href\endcsname\relax
  \def\href#1#2{#2} \def\path#1{#1}\fi

\bibitem{Watts:1998}
D.~J. Watts, S.~H. Strogatz, Collective dynamics of 'small-world' networks,
  Nature 393~(6684) (1998) 440--2.

\bibitem{Barabasi:1999}
A.~L. Barabasi, R.~Albert, Emergence of scaling in random networks, Science
  286~(5439) (1999) 509--12.

\bibitem{Bak:1987}
P.~Bak, C.~Tang, K.~Wiesenfeld, Self-organized criticality: An explanation of
  the 1/f noise, Phys Rev Lett 59~(4) (1987) 381--384.

\bibitem{Jeong:2000}
H.~Jeong, B.~Tombor, R.~Albert, Z.~N. Oltvai, A.~L. Barabasi, The large-scale
  organization of metabolic networks, Nature 407~(6804) (2000) 651--4.

\bibitem{Albert:2000}
R.~Albert, H.~Jeong, A.~L. Barabasi, Error and attack tolerance of complex
  networks, Nature 406~(6794) (2000) 378--82.

\bibitem{Barabasi:2003}
A.~L. Barabasi, E.~Bonabeau, Scale-free networks, Sci Am 288~(5) (2003) 60--9.

\bibitem{Amaral:2000}
L.~A. Amaral, A.~Scala, M.~Barthelemy, H.~E. Stanley, Classes of small-world
  networks, Proc Natl Acad Sci U S A 97~(21) (2000) 11149--52.

\bibitem{Keller:2005}
E.~F. Keller, Revisiting "scale-free" networks, Bioessays 27~(10) (2005)
  1060--8.

\bibitem{Gisiger:2001}
T.~Gisiger, Scale invariance in biology: coincidence or footprint of a
  universal mechanism?, Biol Rev Camb Philos Soc 76~(2) (2001) 161--209.

\bibitem{Clauset:2008}
A.~Clauset, C.~Shalizi, M.~E.~J. Newman, Power-law distributions in empirical
  data, arXiv:0706.1062v1.

\bibitem{Faloutsos:1999}
M.~Faloutsos, P.~Faloutsos, C.~Faloutsos, On power-law relationships of the
  internet topology, in: SIGCOMM 1999, Cambridge, MA, 1999, pp. 251--262.

\bibitem{Mitzenmacher:2006}
M.~Mitzenmacher, Editorial: The future of power law research, Internet
  Mathematics 2~(4) (2006) 525--534.

\bibitem{Newman:2005}
M.~E.~J. Newman, Power laws, pareto distributions and zipf's law, Contemporary
  Physics 46 (2005) 323--351.

\bibitem{Mossa:2002}
S.~Mossa, M.~Barth\'el\'emy, H.~E. Stanley, L.~A. Nunes~Amaral, Truncation of
  power law behavior in ``scale-free'' network models due to information
  filtering, Phys. Rev. Lett. 88~(13) (2002) 138701.

\bibitem{Carvalho:2009}
R.~Carvalho, L.~Buzna, F.~Bono, E.~Gutierrez, W.~Just, D.~Arrowsmith,
  Robustness of trans-european gas networks, Phys Rev E Stat Nonlin Soft Matter
  Phys 80~(1 Pt 2) (2009) 016106.

\bibitem{Sole:2008}
R.~V. Sol\'e, M.~Rosas-Casals, B.~Corominas-Murtra, S.~Valverde, Robustness of
  the european power grids under intentional attack, Phys. Rev. E 77~(2) (2008)
  026102.

\bibitem{Friedkin:1981}
N.~E. Friedkin, The development of structure in random networks: an analysis of
  the effects of increasing network density on five measures of structure,
  Social Networks 3 (1981) 41--52.

\bibitem{Duch:2005}
J.~Duch, A.~Arenas, Community identification using extrmal optimization, Phys
  Rev E 72 (2005) 027104.

\bibitem{Humphries:2008}
M.~D. Humphries, K.~Gurney, Network 'small-world-ness': A quantitative method
  for determining canonical network equivalence, PLoS ONE 3~(4) (2008)
  e0002051.

\bibitem{Kaiser:2006}
M.~Kaiser, C.~C. Hilgetag, Nonoptimal component placement, but short processing
  paths, due to long-distance projections in neural systems, PLoS Comput Biol 2
  (2006) e95.

\bibitem{Wagner:2001}
A.~Wagner, D.~A. Fell, The small world inside large metabolic networks, Proc
  Biol Sci 268 (2001) 1803--1810.

\bibitem{Newman:2003b}
M.~E.~J. Newman, The structure and function of complex networks, SIAM Review 45
  (2003) 167--256.

\bibitem{Martinez:1991}
N.~D. Martinez, Artifacts or attributes? effects of resolution on the little
  rock lake food web, Ecol Monogr 61 (1991) 367--392.

\bibitem{Achard:2006}
S.~Achard, R.~Salvador, B.~Whitcher, J.~Suckling, E.~Bullmore, A resilient,
  low-frequency, small-world human brain functional network with highly
  connected association cortical hubs, J Neurosci 26 (2006) 63--72.

\bibitem{Huxham:1996}
M.~Huxham, S.~Beaney, D.~Raffaelli, Do parasites reduce the chances of
  triangulation in a real food web?, Oikos 76 (1996) 284--300.

\bibitem{Jeong:2001}
H.~Jeong, S.~P. Mason, A.~L. Barabasi, Z.~N. Oltvai, Lethality and centrality
  in protein networks, Nature 411 (2001) 41--42.

\bibitem{Knuth:1993}
D.~E. Knuth, The Stanford GraphBase: A Platform for Combinatorial Computing,
  Addison-Wesley, New York, 1993.

\bibitem{Newman:2006a}
M.~E.~J. Newman, Finding community structure in networks using the eigenvectors
  of matrices, Phys Rev E 74 (2006) 036104.

\bibitem{Albert:1999}
R.~Albert, H.~Jeong, A.~L. Barabasi, Diameter of the world-wide web, Nature 401
  (1999) 130--131.

\bibitem{Newman:2001a}
M.~E.~J. Newman, The structure of scientific collaboration networks, Proc Natl
  Acad Sci USA 98 (2001) 404--409.

\bibitem{Newman:2001b}
M.~E.~J. Newman, S.~H. Strogatz, D.~J. Watts, Random graphs with arbitrary
  degree distributions and their applications, Phys Rev E 64 (2001) 026118.

\bibitem{Lusseau:2003}
D.~Lusseau, The emergent properties of a dolphin social network, Proc Biol Sci
  270 (2003) S186--S188.

\bibitem{Guimera:2003}
R.~Guimera, L.~Danon, A.~Diaz-Guilera, F.~Giralt, A.~Arenas, Self-similar
  community structure in a network of human interactions, Phys Rev E 68 (2003)
  065103(R).

\bibitem{Ebel:2002}
H.~Ebel, L.~I. Mielsch, S.~Bornholdt, Scale-free topology of e-mail networks,
  Phys Rev E 66 (2002) 035103.

\bibitem{Newman:2002}
M.~E.~J. Newman, S.~Forrest, J.~Balthrop, Email networks and the spread of
  computer viruses, Phys Rev E 66 (2002) 035101.

\bibitem{Girvan:2002}
M.~Girvan, M.~E.~J. Newman, Community structure in social and biological
  networks, Proc Natl Acad Sci USA 99 (2002) 7821--7826.

\bibitem{Conyon:2006}
M.~J. Conyon, M.~R. Muldoon, The small world of corporate boards, J Bus Finan
  Account 33 (2006) 1321--1343.

\bibitem{Gleiser:2003}
P.~M. Gleiser, L.~Danon, Community structure in jazz, arXiv:cond-mat (2003)
  0307434v2.

\bibitem{Zachary:1977}
W.~Zachary, An information flow model for conflict and fission in small groups,
  Journal of Anthropological Research 33 (1977) 452--473.

\bibitem{Castro:1999}
R.~d. Castro, J.~W. Grossman, Famous trails to paul erdos, Math Intelligencer
  21 (1999) 51--63.

\bibitem{Ozgur:2004}
A.~Ozgur, H.~Bingol, Social network of co-occurrence in news articles, Computer
  and Information Sciences - ISCIS LNCS3280 (2004) 688--695.

\bibitem{Bearman:2004}
P.~S. Bearman, J.~Moody, K.~Stovel, Chains of affection: The structure of
  adolescent romantic and sexual networks, Am J Sociol 110 (2004) 44--91.

\bibitem{Cancho:2001}
R.~F.~i. Cancho, C.~Janssen, R.~V. Sol\'e, Topology of technology graphs: Small
  world patterns in electronic circuits, Phys. Rev. E 64~(4) (2001) 046119.

\bibitem{Newman:2006b}
M.~E.~J. Newman, Unpublished network data available from:
  http://www-personal.umich.edu/$\sim$mejn/netdata/ (2006).

\bibitem{Adamic:2001}
L.~A. Adamic, R.~M. Lukose, A.~R. Puniyani, B.~A. Huberman, Search in power-law
  networks, Phys Rev E 64 (2001) 046135.

\bibitem{Valverde:2002}
S.~Valverde, R.~F. Cancho, R.~V. Sole, Scale-free networks from optimal design,
  Europhys Lett 60 (2002) 512--517.

\bibitem{Newman:2003a}
M.~E.~J. Newman, Mixing patterns in networks, Phys Rev E 67 (2003) 026126.

\bibitem{Sen:2003}
P.~Sen, S.~Dasgupta, A.~Chatterjee, P.~A. Sreeram, G.~Mukherjee, S.~S. Manna,
  Small-world properties of the indian railway network, Phys Rev E 67 (2003)
  036106.

\bibitem{Batagelj:2006}
V.~Batagelj, A.~Mrvar, Pajek datasets available from
  http://vlado.fmf.uni-lj.si/pub/networks/data/ (2006).

\bibitem{Barabasi:1999a}
A.~L. Barabasi, R.~Albert, H.~Jeong, Mean-field theory for scale-free random
  networks, Physica A 272 (1999) 173--187.

\bibitem{Strogatz:2001}
S.~H. Strogatz, Exploring complex networks, Nature 410 (2001) 268--276.

\bibitem{Newman:2002a}
M.~E.~J. Newman, D.~J. Watts, S.~H. Strogatz, Random graph models of social
  networks, Proc Natl Acad Sci USA 99 (2002) 2566--2572.

\bibitem{Bullmore:2009}
E.~Bullmore, O.~Sporns, Complex brain networks: graph theoretical analysis of
  structural and functional systems, Nature Review Neuroscience 10~(3) (2009)
  186--98.

\bibitem{Hayasaka:2010}
S.~Hayasaka, P.~J. Laurienti, Comparison of characteristics between region-and
  voxel-based network analyses in resting-state fmri data, Neuroimage 50~(2)
  (2010) 499--508.

\end{thebibliography}

\end{document}